\documentclass[aps,prb,twocolumn,superscriptaddress]{revtex4}
\usepackage{amsmath,amssymb}
\usepackage{wasysym}
\usepackage{graphicx}
\usepackage{epstopdf}
\usepackage{latexsym}
\usepackage{subfigure}
\usepackage{color}

\newcommand{\bw}{\begin{widetext}}
\newcommand{\ew}{\end{widetext}}
\newcommand{\ba}{\begin{array}}
\newcommand{\ea}{\end{array}}

\newcommand{\bfr}{{\bf r}}
\newcommand{\bfe}{{\bf e}}
\newcommand{\sfS}{{\sf S}}
\newcommand{\sfs}{{\sf s}}
\newcommand{\bfk}{{\bf k}}
\newcommand{\nn}{\nonumber}

\begin{document}
\title{Generic quantum spin ice}
\author{SungBin Lee}
\affiliation{Department of Physics, University of California, Santa Barbara, CA-93106-9530}
\author{Shigeki Onoda}
\affiliation{Condensed Matter Theory Laboratory, RIKEN, 2-1, Hirosawa, Wako 351-0198, Saitama, Japan}
\author{Leon Balents}
\affiliation{Kavli Institute for Theoretical Physics, University of
  California, Santa Barbara, CA-93106-9530}

\date{\today}
\begin{abstract}
  We consider possible exotic ground states of quantum spin ice as realized in rare earth pyrochlores.  Prior work in Ref.\onlinecite{savary:11} introduced a gauge mean field theory (gMFT) to treat spin or pseudospin Hamiltonians for such systems, reformulated as a problem of bosonic spinons coupled to a U(1) gauge field.  We extend gMFT to treat the {\sl most general}, nearest neighbor exchange Hamiltonian, which contains a further exchange interaction, not considered previously.  This term leads to {\sl interactions} between spinons, and requires a significant extension of gMFT, which we provide.  As an application, we focus especially on the non-Kramers materials Pr$_2TM_2$O$_7$ ($TM=$Sn, Zr, Hf, and Ir), for which the additional term is especially important, but for which an Ising-planar exchange coupling discussed previously is forbidden by time-reversal symmetry.  In this case, when the planar XY exchange is unfrustrated, we perform a full analysis and find three quantum ground states: a U(1) quantum spin liquid (QSL), an antiferro-quadrupolar ordered state and a non-coplanar ferro-quadrupolar ordered one.  We also consider the case of frustrated XY exchange, and find that it favors a $\pi$-flux QSL, with an emergent line degeneracy of low energy spinon excitations.  This feature greatly enhances the stability of the QSL with respect to classical ordering.  \end{abstract}

\maketitle

\section{Introduction}
The quest for quantum spin liquid (QSL) ground states, exotic phases of
matter with emergent gauge structure and quasiparticles carrying
fractional quantum numbers\cite{balentsNature}, is an ongoing endeavour
in condensed-matter physics.  Well studied candidates include some
two-dimensional organic crystals\cite{Kanoda} and some inorganic
kagom\'e systems such as herbertsmithite\cite{YoungLee}.  Among
three-dimensional materials, experimental candidates include several
magnetic pyrochlore oxides~\cite{gardner:10} and hyperkagome-lattice
magnets~\cite{okamoto:07}.  Classical spin liquids have been realized in
the spin ices~\cite{bramwell:01}, in which the spins reside on a
pyrochlore lattice and interact via a dominant classical Ising coupling.
It has been shown theoretically that a weak quantum-mechanical
perturbation does {\sl not} produce long-range order in the ground
state~\cite{hermele:04}.  Instead, it lifts the
macroscopic degeneracy of the spin-ice manifold, leaving gapless ``photon''
excitations, describable by an emergent U(1) gauge field.   The photon
exists in a so-called Coulomb phase or U(1) spin liquid phase, which is
stable to all weak perturbations, at zero temperature.

To describe the low-energy physics of magnetic pyrochlore oxides
associated with local magnetic doublets of rare-earth ions, a minimal
pseudospin-$1/2$ model can be introduced on symmetry
grounds\cite{ross:11} (see Eq.~\eqref{eq:1}).  It has also been derived
micropcopically using superexchange theory for various
materials~\cite{onoda:10,onoda:11,onoda:11b}.  This model successfully
explains spin correlations experimentally observed in Yb$_2$Ti$_2$O$_7$
(Refs.~\onlinecite{ross:11,chang:11}).  As can be seen from the general form
of Eq.~\eqref{eq:1}, these comparisons between theory and experiment
also reveal that putative continuous rotational symmetry of the
pseudospins is broken by a significant level of magnetic anisotropy.
Moreover, at least for Yb$_2$Ti$_2$O$_7$ and possibly for other
materials, the Ising interaction remains dominant, in which case the
physics is that of a quantum variant of spin ice~\cite{molavian:07}.  At
a phenomenological level, recent experimental findings suggest the
relevance of the Coulomb phase physics in real rare-earth magnetic
pyrochlore oxides~\cite{ross:11,chang:11,nakatsuji:11}.

Based on this observation, detailed analyses of the {\sl
  non-perturbative} stability of the Coulomb phase and the possible
existence of other phases and phase transitions are called for.  It
must be noted that this is a very complex problem; the general
pseudospin  Hamiltonian in Eq.~\eqref{eq:1} contains four exchange
constants: the Ising exchange $J_{zz}$, and three ``quantum'' terms
$J_\pm$, $J_{z\pm}$, and $J_{\pm\pm}$.  Assuming we start from the
classically frustrated spin ice case $J_{zz}>0$, one then has three
dimensionless couplings $J_\pm/J_{zz}$, $J_{z\pm}/J_{zz}$ and
$J_{\pm\pm}/J_{zz}$, forming a three-dimensional phase space even at
zero temperature.  The development of a comprehensive theory of this
full 3d phase space is a challenging task.

A method for analysis of this problem was developed in
Ref.~\onlinecite{savary:11}, based on a gauge theory reformulation of the
problem on a dual diamond lattice.  There, the original Hamiltonian was
re-expressed as a problem of bosonic {\sl spinons} hopping in the
background of a fluctuating compact $U(1)$ gauge field.  This problem
was in turn subsequently approximated using a mean-field theory.  In
that work, this gauge Mean Field Theory (gMFT) was applied to the corner
of the phase diagram approximately appropriate to Yb$_2$Ti$_2$O$_7$,
with, in our (and their) notation, $J_{\pm\pm}=0$, and $J_\pm>0$.  Both
the expected U(1) QSL phase and an additional exotic state, a Coulomb
ferromagnet, were found, though somewhat limited in their domain of
stability.

Here we extend the theoretical formalism to allow to fully treat the
most generic nearest-neighbor pseudospin-$1/2$ Hamiltonian, i.e. the
fully general form of Eq.~\eqref{eq:1}.  This requires some significant
technical extensions to the analysis in Ref.~\onlinecite{savary:11}.  In
particular, the term $J_{\pm\pm}$ induces {\sl interactions} amongst the
spinons, which may induce pairing and other effects.  Furthermore, in
the case $J_\pm<0$, a non-zero average gauge flux is present and
complicates the dispersion of the spinons.  Surmounting these technical
obstacles, we then apply the extended method to the particular case
where the lowest crystal-field levels of the rare-earth ions are given
by non-Kramers magnetic doublets with integer spins.  This has a direct
relevance to Pr$_2TM_2$O$_7$ with transition-metal elements
$TM=$~Sn~\cite{matsuhira:02,zhou:08},
Zr~\cite{matsuhira:09,nakatsuji:11}, Hf, and
Ir~\cite{matsuhira:02,nakatsuji:06,machida:09}.  One key result of this
analysis is that the U(1) QSL phase is much more stable than in the case
of a Kramers (half-integer spin) doublet, because of the absence of the
coupling between the Ising and planar components of the rare-earth
moments (Sec.~\ref{sec:mf}).  Moreover, the QSL becomes
particularly robust when the U(1)-symmetric planar pseudospin-exchange
interaction is frustrated ($J_\pm<0$).   

The remainder of the paper is organized as follows. In Sec.~\ref{sec:model}, we introduce the most generic nearest-neighbor pseudospin-1/2 Hamiltonian in rare earth pyrochlores and reformulate it as a problem of bosonic spinons coupled to a U(1) gauge field.  Focusing on non-Kramers doublet, we show in Sec.~\ref{sec:mf} a mean-field analysis of the gauge theory, following and extending Ref.~\onlinecite{savary:11}. Within mean-field analysis, we also discuss that the different flux patterns of gauge fields are favored in the presence of (un)frustrated planar XY exchange cases, based on a perturbation argument. In Sec.~\ref{sec:zero}, unfrustrated planar XY exchange case (FM case) is studied as following orders: In Sec.~\ref{subsec:order-parameters} and Sec.~\ref{subsec:self-con}, we discuss the possible order parameters for spinons and the details of self-consistent equations. Then gMFT phase diagram is shown in Sec~\ref{subsec:gMFT} with the characteristics of their phase transitions described in Sec.~\ref{subsec:phase-transitions}. In Sec~\ref{sec:pi}, we discuss the enhancement of the stability of the QSL for the frustrated planar XY-exchange case (AF case). We finally conclude in Sec.~\ref{sec:summary} with a summary of results, a discussion of relevant experiments and open questions.

\section{Compact Abelian Higgs model for pseudospin-$1/2$ quantum spin ice model}
\label{sec:model}

\subsection{Nearest-neighbor pseudospin-$1/2$ quantum spin ice model}
\label{subsec:spin-model}

We start with the generic nearest-neighbor pseudospin-$1/2$ model for quantum spin ice~\cite{onoda:10,onoda:11,onoda:11b,ross:11};
\begin{eqnarray}
H &=& \sum_{\langle i j \rangle}  [ J_{zz} {\sf S}^z_i {\sf S}^z_j - {J_{\pm}} ( {\sf S}^+_i {\sf S}^-_j + {\sf S}^-_i {\sf S}^+_j )
 \nonumber\\
&&+ {J_{\pm \pm}} (\gamma_{ij}  {\sf S}^+_i {\sf S}^+_j + \gamma_{ij}^{*}  {\sf S}^-_i {\sf S}^-_j )
\nonumber\\
 &&+J_{z\pm}({\sf S}^z_i(\zeta_{ij}{\sf S}^+_j+\zeta_{ij}^*{\sf S}^-_j)
 +(\zeta_{ij}{\sf S}^+_i+\zeta_{ij}^*{\sf S}^-_i){\sf S}^z_j)
 ]. 
 \nonumber\\
 \label{eq:1}
\end{eqnarray}
Here, $\sum_{\langle ij \rangle}$ indicates the summation over all the nearest-neighbor sites $i$ and $j$ on the pyrochlore lattice. ${\sf S}_i$ is a (pseudo-)spin $1/2$ operator acting within the Hilbert space of the local doublet of the rare-earth ion located at that site.  Its physical meaning is discussed further below.  The matrix $\gamma_{ij} = 1, e^{ \pm i 2\pi/3}$ (and $\zeta = - \gamma^*$) depends on the bond direction between the neighboring sites $i$ and $j$ as seen in Fig.\ref{fig:1}:
 \begin{eqnarray}
 \gamma_{\mu \nu} = \Big\{
 \begin{array}{ccc}
  \gamma_1  = &1 , &  \bfe_\mu- \bfe_\nu  \in yz \text{  plane  } \nn\\
  \gamma_2  = & e^{i 2\pi/3} , & \bfe_\mu- \bfe_\nu \in xz \text{  plane  } \nn \\
  \gamma_3  =  & e^{-i 2\pi/3} , &\bfe_\mu- \bfe_\nu  \in xy \text{  plane  } 
\end{array}.
\label{eq:2}
\\
 \end{eqnarray}
Here we defined the four vectors, $\bfe_\mu$ ($\mu=0,1,2,3$), connecting a site on the diamond sublattice $A$ to its neighbors.  In our coordinates, where the conventional cubic unit cell is taken of length $1$, these are related to the local $z$ quantization (Ising) axis for the pseudospins given in Table~\ref{tab:2} by $\bfe_\mu = \frac{\sqrt{3}}{4} \hat{z}_\mu$.  

Now we review the physical meaning of the pseudospins.  The $z$
component ${\sf S}^z_i$ is directly proportional to the physical
magnetic moment along the local Ising $\langle111\rangle$ axis.  The
interpretation of the ``in-plane'' pseudospin operators, ${\sf S}_i^x,
{\sf S}_i^y$, differs, however, for the Kramers and non-Kramers cases.
In the former, this is indeed proportional to the magnetic dipole moment
normal to the local $\langle 111\rangle$ axis.   Because in the
non-Kramers case, the in-plane components of the pseudospin are
time-reversal invariant, they must be identified not with the magnetic
dipole moment but for instance the quadrupole moment
${\sf S}_i^\pm \sim
    \{J^\pm_i,J^z_i\}$ in the Pr cases~\cite{onoda:11,onoda:11b}.

This is summarized formally as follows.  Defining local coordinate axes
$\hat{\boldsymbol x}_i$, $\hat{\boldsymbol y}_i$, and $\hat{\boldsymbol z}_i$, as in Table.\ref{tab:2}, the
magnetic dipole moment is given as
\begin{equation} 
\langle {\boldsymbol \mu}_i \rangle =g_{xy}\mu_B(\langle \sfS_i^x \rangle
  \hat{\boldsymbol x}_i + \langle \sfS_i^y \rangle \hat{\boldsymbol y}_i) +g_z\mu_B\langle
  \sfS_i^z \rangle \hat{\boldsymbol z}_i,
\label{eq:3} 
\end{equation}
where for the non-Kramers magnetic doublets which are of our particular
interest in this paper, $g_{xy}=0$.  For this case (and specifically for
Pr$^{3+}$) the quadupole moment
$(\overleftrightarrow{Q}_i)^{\mu\nu}=\mu_B^2\{J^\mu_i,J^\nu_i\}$ is
\begin{equation} 
\langle \overleftrightarrow{Q}_i\rangle\sim
\left(\begin{array}{ccc}
  0 & 0 & \langle \sfS_i^x \rangle \\
  0 & 0 & \langle \sfS_i^y \rangle \\
  \langle \sfS_i^x \rangle & \langle \sfS_i^y \rangle & 0
\end{array}\right).
\label{eq:4} \end{equation}

\begin{table} \centering \label{tab:2} \begin{tabular}{|c| c| c| c| c|} \hline
    $i$ & 0 & 1& 2& 3 \\
    \hline \hline $\hat{x}_i$ & $\frac{1}{\sqrt{2}} (0 1\bar{1} )$ & $\frac{1}{\sqrt{2}} (0 \bar{1} 1)$
    & $\frac{1}{\sqrt{2}} (011)$ &  $\frac{1}{\sqrt{2}} (0 \bar{1} \bar{1} )$ \\
    $\hat{y}_i$ & $\frac{1}{\sqrt{6}} (\bar{2} 11 )$ & $\frac{1}{\sqrt{6}} (\bar{2} \bar{1} \bar{1})$
    & $\frac{1}{\sqrt{6}} (2 1 \bar{1} )$ &  $\frac{1}{\sqrt{6}} (2 \bar{1} 1 )$ \\
    $\hat{z}_i$ & $\frac{1}{\sqrt{3}} (111)$ & $\frac{1}{\sqrt{3}} (1 \bar{1} \bar{1})$
    & $\frac{1}{\sqrt{3}} (\bar{1} 1 \bar{1})$ &  $\frac{1}{\sqrt{3}} (\bar{1} \bar{1} 1)$\\
    \hline \end{tabular} \caption{Local coordinate frames for the four sublattices on the pyrochlore lattice.}  \end{table}

\subsection{Interacting bosonic spinons coupled to compact U(1) gauge fields on a diamond lattice}
\label{subsec:bosonic-spinons}
%
\begin{figure}[t]
\vskip0.5cm
  \includegraphics[width=0.4\textwidth]{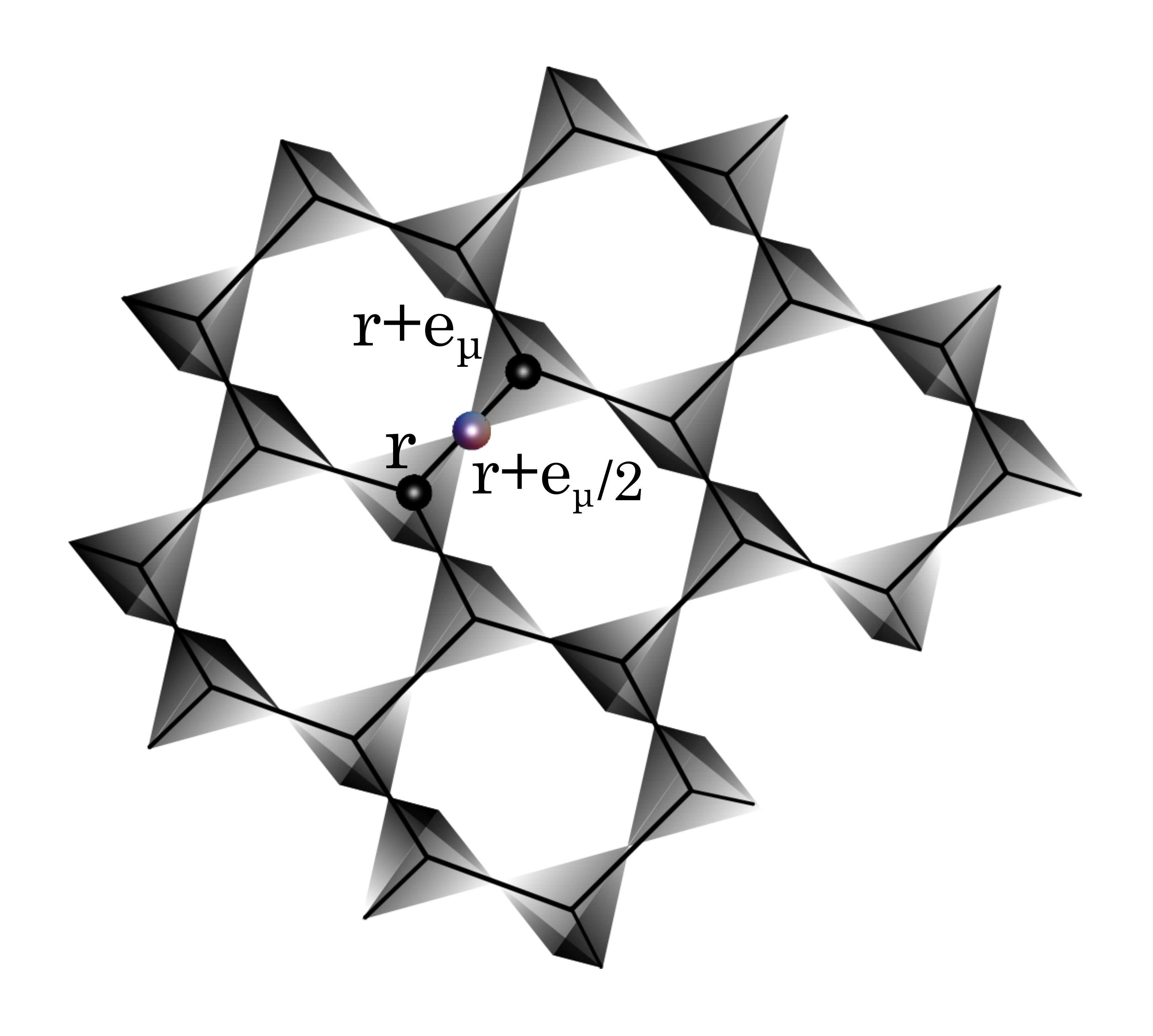}
  \caption{Mapping onto U(1) gauge theory : spinons are defined on diamond lattice sites (black sphere) and gauge fields live on their links (black solid line). $J_\pm$ ($J_{\pm \pm}$) shows quadratic (quartic) spinon hopping process in Eq.\eqref{eq:11}.}
\label{fig:1} 
\end{figure}
%

In this section, we {\sl exactly} re-express the Hamiltonian
Eq.~\eqref{eq:1} as a $U(1)$ gauge theory on the dual diamond lattice.
The formulation is identical, apart from some minor notational changes,
to the one in Ref.~\onlinecite{savary:11}, except that we now include
the $J_{\pm\pm}$ term.

First, the Ising interaction term is simply expressed as $J_{zz}\sum_\bfr Q_\bfr^2$ with a gauge (``electric'') charge~\cite{hermele:04,isakov:04}
\begin{equation}
Q_\bfr = \eta_\bfr \sum_{\mu=0}^3 \sfS^z_{\bfr + \eta_\bfr \bfe_\mu/2},
 \label{eq:5}
\end{equation}
where the coordinate $\bfr$ labels the centers of the pyrochlore
tetrahedra, which form the ``dual'' diamond lattice.  We furthermore
defined $\eta_\bfr =1$ and -1, when $\bfr$ is on the $A$ and $B$ diamond
sublattice, respectively. (see Fig.\ref{fig:1}) Note that the electric
charge $Q_\bfr$ takes integer values. Then, the corresponding ``electric
field'' can be taken as a directed link variable,
\begin{equation}
  E_{\bfr,\bfr+\eta_\bfr\bfe_\mu}=\eta_\bfr{\sf S}_{\bfr+\eta_\bfr\bfe_\mu/2}^z=\eta_\bfr{\sf s}^z_{\bfr,\bfr+\eta_\bfr\bfe_\mu}.
  \label{eq:6}
\end{equation}
With these definitions, Eq.~\eqref{eq:5} may be regarded as the lattice
analog of Gauss' law.  Maintaining the constraint in Eq.~\eqref{eq:5},
one then represents the  planar components ${\sf S}^\pm$ of the
pseudospin operator as
\begin{equation}
{\sf S}_{\bfr+\bfe_\mu/2}^+ = \Phi_{\bfr}^\dagger {\sf s}^+_{\bfr , \bfr
  + \bfe_\mu} \Phi_{\bfr + \bfe_\mu} \qquad \textrm{for} \qquad \bfr \in A,
\label{eq:7}
\end{equation}
where $\Phi_\bfr$ and $\Phi^\dagger_\bfr$ are annihilation and creation operators of bosonic spinons that decrease and increase the ``electric charge'' $Q_\bfr$, respectively,
\begin{equation}
  [\Phi_\bfr,Q_\bfr]=\Phi_\bfr,\ \ \ 
  [\Phi_\bfr^\dagger,Q_\bfr]=-\Phi_\bfr^\dagger.
  \label{eq:8}
\end{equation}
Note that ${\sf s}^\pm$ conceptually plays the role of the exponential
of the gauge vector potential $A$, i.e. an element of the gauge group,
which creates or annihilates electric field quanta.  There is, however,
no utility in explicitly introducing $E$ and $A$ operators, at least at
the mean field level we proceed with in the following, and we will not
do so.  

It is convenient to introduce a rotor variable $\varphi_\bfr$ canonical
conjugate to $Q_\bfr$;
\begin{equation}
[ \varphi_\bfr, Q_\bfr ] = i,
\label{eq:9}
\end{equation}
which gives
\begin{equation}
\Phi_\bfr = e^{-i \varphi_\bfr}, \ \ \ \Phi_\bfr^\dagger \Phi_\bfr = 1.
\label{eq:10}
\end{equation}
Using the representation in Eqs.~\eqref{eq:5}-~\eqref{eq:7}, we rewrite spin Hamiltonian Eq.~\eqref{eq:1} 
\bw
\begin{eqnarray}
H_{QED} &=& 
  \frac{J_{zz}}{2} \sum_{\bfr} Q_\bfr^2 
 -{J_{\pm}} \sum_{\bfr} \sum_{ \mu \neq \nu}  \Phi_{\bfr+ \eta_\bfr \bfe_\mu}^\dagger  \Phi_{ \bfr+ \eta_\bfr \bfe_\nu}  \sfs_{\bfr,\bfr+ \eta_\bfr \bfe_\mu}^{-\eta_\bfr} \sfs_{\bfr, \bfr+ \eta_\bfr \bfe_\nu}^{+\eta_\bfr}
 \nn \\
&&  + \frac{J_{\pm \pm}}{2}  \sum_{\bfr} \sum_{\mu \neq \nu}  
( \gamma_{\mu \nu}^{-2\eta_\bfr} \Phi_{\bfr }^\dagger \Phi_{\bfr }^\dagger 
 \Phi_{\bfr + \eta_\bfr \bfe_\mu} \Phi_{\bfr + \eta_\bfr \bfe_\nu} 
  \sfs_{\bfr , \bfr+ \eta_\bfr \bfe_\mu}^{\eta_\bfr}  \sfs^{\eta_\bfr}_{\bfr, \bfr+ \eta_\bfr \bfe_\nu} + h.c )
  \nn \\
  &&-J_{z\pm}\sum_{\bfr}\sum_{\mu\ne\nu}{\sf s}^z_{\bfr, \bfr + \eta_\bfr \bfe_\mu}\left(\gamma_{\mu\nu}^{-\eta_\bfr}\Phi_\bfr^\dagger\Phi_{\bfr + \eta_\bfr \bfe_\nu}\sfs^{\eta_\bfr}_{\bfr, \bfr+ \eta_\bfr \bfe_\nu}+h.c.\right)
  + \text{const..}
 \label{eq:11}
 \end{eqnarray}
 \ew
The total ``electric charge'' $Q=\sum_{\bfr}Q_\bfr$ commutes with the Hamiltonian $H_{QED}$. Furthermore, $H_{QED}$ is invariant under the gauge transformation,
\begin{eqnarray}
\Big\{ 
\begin{array}{cc}
\Phi_\bfr \rightarrow &  \Phi_\bfr e^{ - i\chi_\bfr} \\
\sfs^\pm_{\bfr \bfr'} \rightarrow & \sfs^\pm_{\bfr \bfr'} e^{ \pm i ( \chi_{\bfr'} - \chi_{\bfr} )}.
\end{array}
\label{eq:12}
\end{eqnarray}
This completes the reformulation as a compact U(1) gauge theory with
bosonic matter: a compact interacting Abelian Higgs model.
 
\section{Mean-field theory}
\label{sec:mf}
  
We now proceed with a mean-field analysis of the gauge theory in
Eq.~\eqref{eq:11}, following and extending Ref.~\onlinecite{savary:11}.
To distinguish this from ordinary Curie-Weiss mean field theory, we
denote this treatment as gauge Mean Field Theory (gMFT).\cite{savary:11}
Specifically, we decouple the various terms in Eq.~\eqref{eq:11} as
follows: 
\begin{widetext}
\begin{eqnarray}
  s^+_{\bfr,\bfr+\bfe_\mu} s^-_{\bfr,\bfr+\bfe_\nu} \Phi^\dagger_{\bfr+\bfe_\mu} \Phi_{\bfr+\bfe_\nu} 
  &\rightarrow&  
  \langle s^+_{\bfr,\bfr+\bfe_\mu} \rangle \langle s^-_{\bfr,\bfr+\bfe_\nu} \rangle 
  \left( \Phi^\dagger _{\bfr+\bfe_\mu}\Phi_{\bfr+\bfe_\nu} - \langle \Phi^\dagger_{\bfr+\bfe_\mu} \Phi_{\bfr+\bfe_\nu} \rangle \right)
  \nonumber\\
  &&+  \left(s^+_{\bfr,\bfr+\bfe_\mu} \langle s^-_{\bfr,\bfr+\bfe_\nu} \rangle + \langle s^+_{\bfr,\bfr+\bfe_\mu} \rangle s^-_{\bfr,\bfr+\bfe_\nu} - \langle s^+_{\bfr,\bfr+\bfe_\mu} \rangle \langle s^-_{\bfr,\bfr+\bfe_\nu} \rangle \right)
  \langle \Phi^\dagger_{\bfr+\bfe_\mu} \Phi_{\bfr+\bfe_\nu} \rangle,
 \label{eq:13}\\
 s^+_{\bfr,\bfr+\bfe_\mu} s^+_{\bfr,\bfr+\bfe_\nu} \Phi^\dagger_\bfr \Phi^\dagger_\bfr \Phi_{\bfr+\bfe_\mu} \Phi_{\bfr+\bfe_\nu} 
 &\rightarrow &
 \langle s^+_{\bfr,\bfr+\bfe_\mu} \rangle \langle s^+_{\bfr,\bfr+\bfe_\nu} \rangle
 \left[\Phi^\dagger_\bfr \Phi^\dagger_\bfr \langle \Phi_{\bfr+\bfe_\mu} \Phi_{\bfr+\bfe_\nu} \rangle
  +  \langle \Phi^\dagger_\bfr \Phi^\dagger_\bfr \rangle \Phi_{\bfr+\bfe_\mu} \Phi_{\bfr+\bfe_\nu} 
  - 2\langle \Phi^\dagger_\bfr \Phi^\dagger_\bfr \rangle \langle \Phi_{\bfr+\bfe_\mu} \Phi_{\bfr+\bfe_\nu} \rangle
  \right.
  \nonumber\\
  &&\left.
  +2\left(\Phi^\dagger_\bfr \Phi_{\bfr+\bfe_\mu} \langle \Phi^\dagger_\bfr \Phi_{\bfr+\bfe_\nu} \rangle
  +  \langle \Phi^\dagger_\bfr \Phi_{\bfr+\bfe_\mu} \rangle \Phi^\dagger_\bfr \Phi_{\bfr+\bfe_\nu} 
  - 2\langle \Phi^\dagger_\bfr \Phi_{\bfr+\bfe_\mu} \rangle \langle \Phi^\dagger_\bfr \Phi_{\bfr+\bfe_\nu} \rangle\right)\right]
  \nonumber\\
  &&+ \left(\langle \Phi^\dagger_\bfr \Phi^\dagger_\bfr \rangle \langle \Phi_{\bfr+\bfe_\mu} \Phi_{\bfr+\bfe_\nu} \rangle
  +2\langle \Phi^\dagger_\bfr \Phi_{\bfr+\bfe_\mu} \rangle \langle \Phi^\dagger_\bfr \Phi_{\bfr+\bfe_\nu} \rangle\right)
  \nonumber\\
  &&\times \left(s^+_{\bfr,\bfr+\bfe_\mu} \langle s^-_{\bfr,\bfr+\bfe_\nu} \rangle + \langle s^+_{\bfr,\bfr+\bfe_\mu} \rangle s^-_{\bfr,\bfr+\bfe_\nu} - \langle s^+_{\bfr,\bfr+\bfe_\mu} \rangle \langle s^-_{\bfr,\bfr+\bfe_\nu} \rangle \right).
  \label{eq:14}
\end{eqnarray}
\end{widetext}
The second decoupling (Eq.~\eqref{eq:14}) is introduced here (and was
not considered in Ref.~\onlinecite{savary:11}) to deal with the
$J_{\pm\pm}$ interaction, which involves not only interaction between
the ``gauge fields'' (${\sf s}_{\bfr,\bfr'}$) and spinons, but also
between the spinons themselves (being quartic in
$\Phi_\bfr^{\vphantom\dagger},\Phi^\dagger_\bfr$ operators).

After the above decoupling, some ansatz must be made to determine the
form of various expectation values.  Of particular interest are the
``magnetic'' expectation values 
\begin{equation}
  \label{eq:15}
  \langle {\sf s}^\pm_{\bfr,\bfr'}\rangle = |\langle {\sf
    s}^\pm_{\bfr,\bfr'}\rangle| e^{\pm i \overline{A}_{\bfr,\bfr'}},
\end{equation}
which are formally similar to the bond operators appearing in large $N$
slave particle theories of quantum antiferromagnets.  In particular, the
{\sl phase} $\overline{A}_{\bfr,\bfr'}$of this expectation value has the physical
interpretation of a sort of ``average'' gauge field experienced by the
spinons.  Though this is not itself gauge invariant (see
Eq.~\eqref{eq:12}), the net {\sl flux} of this gauge field, or
equivalently the product of $\langle {\sf s}^+_{\bfr,\bfr'}\rangle$
expectation values around any closed loop, is physically meaningful.
Different patterns of this flux describe different QSL states, with
different Projected Symmetry Groups, or PSGs.\cite{wen:02} Within the
gMFT formalism, the choice between different flux patterns, or PSGs,
should be made based on comparison of the mean-field free energy for
different {\sl Ans\"atze}.

Instead, we choose the flux pattern based on a non-mean-field, but
perturbative argument.  This has the advantage of being simpler, and
also correct beyond mean field theory in the perturbative regime, but
could in principle break down by missing some other phase at larger
coupling.  We leave a more exhaustive study of energetics of different
PSGs as an open problem for the future, but expect that the choice made
here is probably correct in most cases of interest.  

\begin{figure}[htb]
\begin{center}
\vskip0.5cm
  \includegraphics[width=0.45\textwidth]{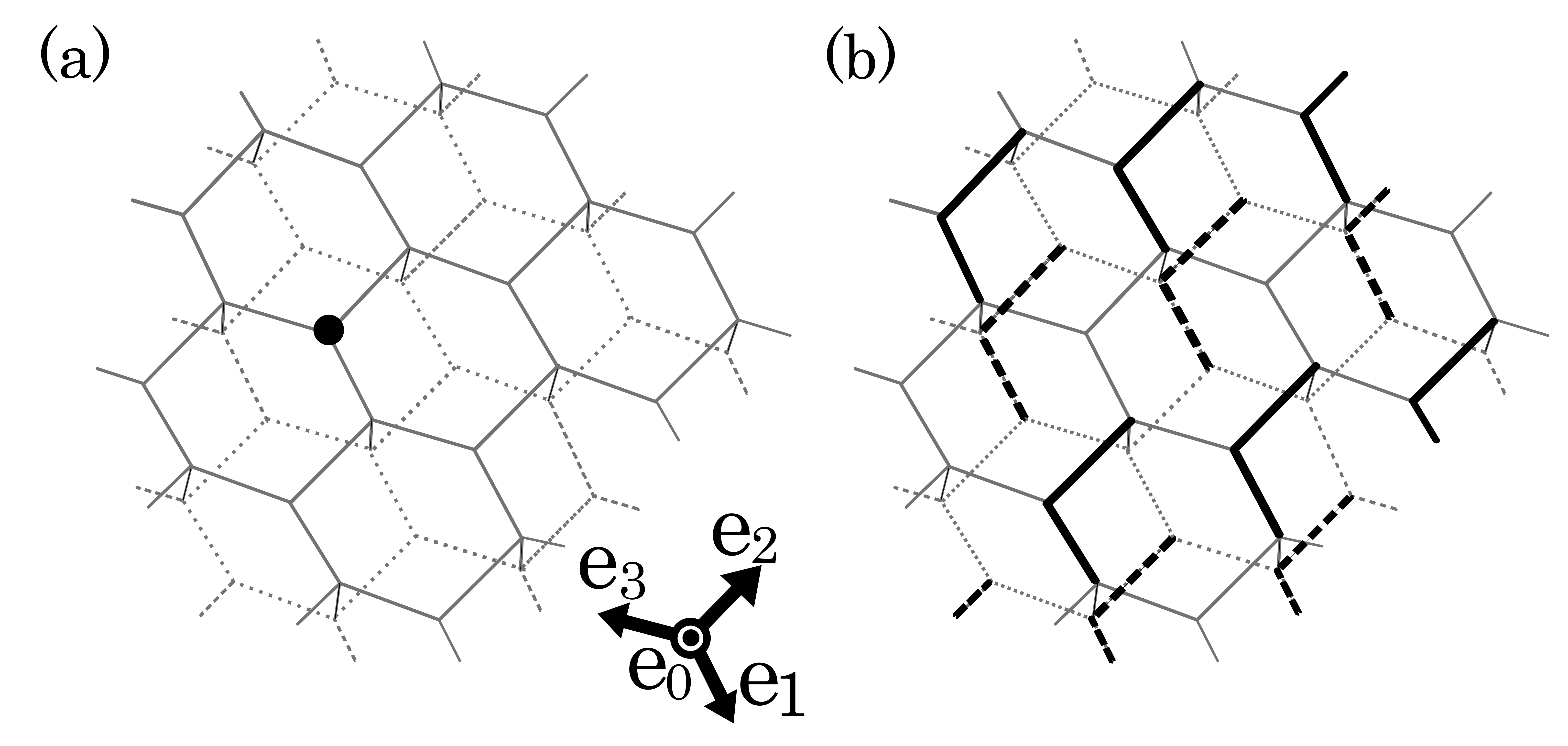}
\end{center}
\caption{Mean-field ansatz for gauge fields that preserve the symmetry
  of Hamiltonian Eq.~\eqref{eq:11}: (a) 0-flux state, and (b)
  $\pi$-flux state. In (b), the thick links have
  $\overline{A}_{\bfr,\bfr'} = \pi$ and other links have $ \overline{
    A}_{\bfr,\bfr'} =0$.
   Thick solid (dashed) links are aligned in upper (lower) hexagonal plane perpendicular to $\langle 111 \rangle$ direction in a diamond lattice. }
\label{fig:2}
\end{figure}

In the perturbative regime, $J_{\pm \pm}/J_{zz} \lesssim J_{\pm} /
J_{zz} \ll 1$, the leading contribution in degenerate perturbation
theory \cite{hermele:04} occurs at order $(J_{\pm}^3 / J_{zz}^2 )$ (the
contribution from $J_{\pm\pm}$ is subdominant), and gives an term in the
Hamiltonian proportional to the sum of the cosine of the flux through
each hexagonal plaquette of the dual diamond lattice,
\begin{equation}
  \label{eq:16}
  H_{\rm ring} \sim - J^3_{\pm} /J^2_{zz} \sum_{\hexagon} \cos (\nabla
  \times \overline{A} ),
\end{equation}
where $(\nabla \times \overline{A})_{\hexagon} \equiv \sum_{i\in\hexagon}
\overline{A}_{\bfr_i,\bfr_{i+1}}$ denotes a lattice curl of the gauge field
$\overline{A}_{\bfr, \bfr'}$, i.e. the flux through the hexagon containing the
sites $i$.  This calculation determines the PSG to be: (1) for
$J_{\pm}>0$, a $0$-flux state, with $\cos (\nabla \times \overline{A} )=1$ or (2)
for $J_\pm<0$, a $\pi$-flux state with $\cos (\nabla \times \overline{A} )=-1$.
This is somewhat consistent with the physical intuition that for
$J_\pm<0$, the XY pseudospin order is frustrated, requiring formation of
a more complex ground state.  Indeed, we will see later that quantum
fluctuations are greatly enhanced in $\pi$-flux state, leading as a
consequence to much enhanced domain of stability of the QSL phase
relative to the case $J_\pm>0$.  In the following, we will consider the
zero-flux and $\pi$-flux cases separately.
Fig.\ref{fig:2} shows mean-field ansatz of gauge fields for zero-flux and $\pi$-flux cases.
Thick links have $\overline{A}_{\bfr, \bfr'} = \pi$ and other links have $\overline{A}_{\bfr, \bfr'} = 0$.
There exist many other gauge field configurations that produce the same physical state.

%
%

\section{FM case ($J_{\pm}>0$) :  zero-flux state}
\label{sec:zero}

In this section, we carry out an analysis of the ``ferromagnetic'' case,
$J_\pm>0$, for which the $0$-flux state is stabilized.  The zero flux
state was considered in Ref.~\onlinecite{savary:11}, but with
$J_{\pm\pm}=0$. 

\subsection{Order parameters}
\label{subsec:order-parameters}

The gMFT treatment introduces several self-consistently determined ``order parameters'', whose interpretation requires additional care in comparison to ordinary mean field theory, owing to gauge redundancy.  We discuss this now.  In the ``gauge'' sector, there are two types of expectation values: $\langle {\sf s}_{\bfr,\bfr'}^z\rangle$, which is gauge invariant and directly proportional to the local magnetic moment, see Eqs.~(\ref{eq:2},\ref{eq:6}), and $\langle {\sf s}_{\bfr,\bfr'}^+\rangle$, which as discussed above is not gauge invariant and related to the ``average'' and fluctuations of the magnetic vector potential.  There are several other order parameters in the ``matter'' sector.  While it is not explicit as a decoupling in the gMFT scheme, the spinon condensate itself, $\langle \Phi_\bfr\rangle$, is an important (non-gauge invariant) order parameter.  In comparison to the prior case, the decomposition in Eq.~\eqref{eq:14} introduces two new types of order parameters.  First, there are pairing terms between two spinon fields on the same sublattice, $\langle \Phi_{\bfr+\eta_\bfr\bfe_\mu}\Phi_{\bfr+\eta_\bfr\bfe_\nu}\rangle$.  This composite field carries electric gauge charge 2.  Second, there are particle-hole terms which are gauge neutral but connect the two sublattices, $\langle \Phi_\bfr^\dagger \Phi\vphantom{\dagger}_{\bfr+\eta_\bfr\bfe_\mu}\rangle$.  The possibility of non-zero expectation values of these fields enlarges the spectrum of phases which may occur within gMFT.

For the FM case, we have found that the mean field solutions do not break translational symmetry.  Presuming this to be true, we can classify the different possible types of solutions by their order parameter expectation values, and we discuss the physical meaning of these patterns now.  The different possibilities are listed in Table.~\ref{tab:1} and summarized below.

\subsubsection{Coulomb phases}

Within a mean field description, Coulomb phases are those in which the U(1) gauge symmetry is unbroken by charged condensates, i.e. $\langle \Phi_\bfr \rangle = \langle \Phi_{\bfr+\eta_\bfr\bfe_\mu}\Phi_{\bfr+\eta_\bfr\bfe_\nu}\rangle =0$, and where the gauge fluctuations are not so strong as to wash out the average of the gauge elements, i.e. $\langle {\sf s}_{\bfr,bfr'}^\pm\rangle \neq 0$.  This leaves $\langle {\sf s}_{\bfr,\bfr'}^z\rangle$ and $\langle \Phi_\bfr \Phi_{\bfr+\eta_\bfr\bfe_\mu}\rangle$ undetermined, and depending upon their values, different phases can be realized.

\paragraph{U(1) QSL ---}

When $\langle {\sf s}_{\bfr,\bfr'}^z\rangle=\langle \Phi_\bfr \Phi_{\bfr+\eta_\bfr\bfe_\mu}\rangle=0$, all physical (global) symmetries of the system are unbroken.  Due to the non-zero  value of $\langle {\sf s}_{\bfr,\bfr'}^\pm\rangle $, the spinons are able to coherently hop and propagate.  Thus this phase may be characterized as a $U(1)$ QSL with propagating spinons and an emergent $U(1)$ gauge field (which appears in the mean field treatment by fluctuations of the phase $\overline{A}_{\bfr,\bfr'}$).  This is the realization in gMFT of the QSL discussed perturbatively in Ref.~\onlinecite{hermele:04}.  

\paragraph{Coulombic ordered phase ---}

When either $\langle \Phi_\bfr^\dagger \Phi_{\bfr'} \rangle \neq 0$, $\langle {\sf s}_{\bfr,\bfr'}^z\rangle \neq 0$, or both, global physical symmetries of the pseudospin Hamiltonian are broken.  For instance, $\langle {\sf s}_{\bfr,\bfr'}^z\rangle \neq 0$ implies time-reversal symmetry breaking and the presence of spontaneous magnetic dipole moments oriented along the local $\langle 111\rangle$ axes.  However, phases of this nature are not trivial ordered states, since like the QSL, they host propagating deconfined spinons and an emergent gapless Coulomb gauge field.   A particular example of this type, dubbed the ``Coulomb ferromagnet'', was discussed in Ref.~\onlinecite{savary:11}, but more general such phases may occur when the full phase space is considered.  They do not, however, occur in the gMFT solution in the following subsection.

\subsubsection{Confined phase --- Ising order}

Another possible phase is one in which the gauge fluctuations are sufficiently strong that they destroy the bond operator expectation values, $\langle {\sf s}^\pm_{\bfr,\bfr'}\rangle =0$.  In this case, the amplitude for spinons to hop vanishes, and so they do not coherently propagate.  This should be considered therefore a {\sl confined} phase.  Since the ${\sf s}_{\bfr,\bfr'}$ are spin-1/2 operators, the only state consistent with the above condition is polarized along the $z$ direction, so that $\langle {\sf s}_{\bfr,\bfr'}^z\rangle \neq 0$, implying broken time-reversal symmetry and Ising magnetic order.  In this state, the gauge fluctuations are insignificant, and there are no emergent low-energy gauge fields.   Indeed, such a state would be considered a classically ordered spin ice~\cite{siddharthan:01,melko:04}.   In our calculations, we find that this phase does not occur in the region of phase space we studied.  

\subsubsection{Higgs phase --- quadrupolar order}

It is well known that in gauge theories there are two ways to remove the gauge fields from the low energy physics: confinement, as described above, and the Higgs mechanism.  In the Higgs mechanism, a boson carrying the fundamental gauge charge condenses, leading to a ``Meissner effect'' for the gauge flux and a gap for the photon.  Though the Higgs mechanism for the {\sl transition} seems very different from that occuring in confinement, it is believed that, in the absence of global symmetries, the confinement and Higgs {\sl phases} are indistinct, and can be adiabatically transformed into one another.  Therefore, like the confined state discussed above, the Higgs phase(s) which occur in gMFT are conventional states of matter, absence any exotic excitations or topological properties.  In our case, however, the Higgs phases can be distinguished by symmetry from the confined one.

To see this,  consider a state with spinon condensation at wavevector $\bfk$, i.e., 
\begin{equation}
\langle \Phi_{\bfr}\rangle=\langle\Phi_\bfk\rangle e^{i\bfk\cdot\bfr}\ne0. 
\label{eq:17}
\end{equation}
For the spinons to condense, they must obviously propagate, so we must
also assume $\langle {\sf s}_{\bfr,\bfr'}^\pm\rangle \neq 0$.  The mean
field result is then long-range ordering of pseudospins on the  XY plane
(which corresponds to quadrupolar ordering in physical terms) since 
\begin{equation}
 \langle \sfS_\bfr^+  \rangle = \langle \Phi^\dagger_\bfr \rangle \langle \sfs^+_{\bfr, \bfr+\bfe_\mu} \rangle \langle \Phi_{\bfr + \bfe_\mu} \rangle \neq 0.
\label{eq:18}
\end{equation}
We indeed find Higgs phases of this type in the solution of the gMFT equations given below.

\subsubsection{Charge 2 Higgs phase --- $Z_2$ QSL}
\label{subsubsec:charge-2-higgs}

The remaining possible phase is one in which the charge 2 Higgs (spinon pair) condensate is non-vanishing, $\langle \Phi_{\bfr +\eta_\bfr \bfe_\mu } \Phi_{\bfr+\eta_\bfr\bfe_\nu}\rangle \neq 0$, but the fundamental spinon is uncondensed, $\langle \Phi_\bfr\rangle =0$.  In this case, the U(1) gauge symmetry is broken down to $Z_2$, and the resulting state is a gapped QSL with only Ising-like gauge charges.  Spinons remain deconfined, but become mixtures of particles and anti-particles, much like Bogoliubov quasiparticles in a superconductor are mixtures of electrons and holes.  Due to the absence of spinon condensation, this state generically has vanishing spin expectation values and need not break symmetries.   Like the confined and Coulombic ordered phases, this state does not, however, occur as a ground state in gMFT, in the parameter regime we have so far studied.

\begin{table}[hbtp]
\centering
\begin{ruledtabular}
\begin{tabular}{c| c| c| c| c|c}
 & $\langle \sfs^z_{\bfr ,\bfr \pm \bfe_\mu}  \rangle $  & $\langle \sfs^\pm_{\bfr ,\bfr \pm \bfe_\mu}  \rangle $  & $\langle \Phi_\bfr \rangle$ & $ \langle \Phi_{\bfr} \Phi_{\bfr} \rangle $ & $\langle \Phi_\bfr^\dagger \Phi_{\bfr\pm\bfe_\mu}\rangle$
\\
\hline
Ising order & $\neq 0$ & 0 & 0 & 0 & 0 \\
(confined) & & & &\\
\hline
QSL & & & & &\\
U(1) & 0 & $\neq 0$ & 0 & 0 & 0 \\
Z$_2$ & 0 & $\neq 0$ & 0 & $\neq 0$ & 0\\
(charge-2 Higgs) & & & &\\
\hline
XY order & & & & &\\
U(1) & 0 & $\neq 0$ & 0 & 0 &$\neq 0$ \\
Classical & 0 & $\neq 0$ & $\neq 0$ & $\neq 0$ & $\neq 0 $ \\
(confined Higgs) & & & &\\
\end{tabular}
\end{ruledtabular}
\caption{Classification of possible phases occuring in the gMFT treatment. } 
\label{tab:1}
\end{table}

\subsection{Self-consistent equations}
\label{subsec:self-con}

Once the replacements in Eqs.~(\ref{eq:13},\ref{eq:14}) have been made, the mean field Hamiltonian reduces to a sum of Hamiltonians,
\begin{equation}
  \label{eq:19}
  H \rightarrow H_{gMFT} = H_{\sf s} + H_{\Phi},
\end{equation}
where $H_{\sf s}$ describes the gauge sector as decoupled ``spins''
${\sf\boldsymbol s}_{\bfr,\bfr'}$, and the spinon part $H_\Phi$
contains only $\Phi_{\bfr}^{\vphantom\dagger},\Phi^\dagger_{\bfr}$ and
$Q_\bfr$ operators and is quadratic in them.  The gauge part $H_{\sf
  s}$ is trivially soluble, as different bonds are decoupled, and each
${\sf s}_{\bfr,\bfr'}$ is then treated simply as a $s=1/2$ spin in a
field.  The spinon part, however, is non-trivial, and actually still
strongly interacting, since $\Phi_\bfr$ is actually defined in terms
of the fundamental rotor field $\varphi_\bfr$, c.f. Eq.~\eqref{eq:10}.
To proceed with it, we follow
Refs.~\onlinecite{FlorensGeorges},\onlinecite{savary:11}, and replace
the ``hard'' constraint on $\Phi_\bfr^\dagger
\Phi_\bfr^{\vphantom\dagger}=1$ with a ``soft'' an average one,
enforced by a Lagrange multiplier $\lambda_\bfr$.  This is equivalent,
quantum mechanically, to promoting the spinon field to a complex
rotor, with $\Phi_\bfr = x_\bfr + i y_\bfr$ and $Q_\bfr = p_{x_\bfr} +
i p_{y_\bfr}$, where $x$ and $p$ variables are canonically conjugate
coordinates and momenta.
After this substitution, $H_\Phi$ becomes quadratic and soluble. The Lagrange
multiplier, appearing as a mass for $\Phi_\bfr$, is adjusted to maintain
$\langle \Phi_\bfr^\dagger \Phi_\bfr^{\vphantom\dagger}\rangle=1$ on
every site.

In the following, we use this formulation to calculate the necessary
expectation values and impose self-consistency.  We assume here a zero
flux state, and no breaking of translational symmetry.  We also neglect
the $J_{z\pm}$ coupling, in which case one can show that the energy is
minimized when $\langle {\sf s}_{\bfr,\bfr'}^z\rangle=0$. Then in
general there are many variables: 4 distinct gauge fields $\langle {\sf
  s}_{\bfr,\bfr'}^+\rangle$ on the four orientations of diamond bonds, 8
spinon pair fields (2 on a single site and 6 distinct orientations of
pairs connecting same-sublattice sites in different unit cells), 4 A-B
sublattice mixing field on diamond bonds, and 2 Lagrange multipliers,
one for each of the two basis sites.  This makes 18 distinct mean field
parameters, a general analysis of which is daunting.  To proceed, we
looked for self-consistent solutions with fewer parameters.  We employed
a rather general ansatz, containing both pairing and A-B sublattice
mixing, but imposing some discrete symmetry constraints.  We discuss
the comparison of the {\sl energy} of these solutions in the subsequent
subsection.

\subsubsection{ Mean-Field Ansatz}
\label{subsec:mf-ansatz}

We introduce the following ansatz including both pairing
and A-B sublattice mixing:
\begin{eqnarray}
\Delta &=& \langle s^\pm_{\bfr, \bfr \pm \bfe_\mu} \rangle, \label{eq:20} \\
\chi_0^A &=& \langle \Phi_{\bfr_A} \Phi_{\bfr_A} \rangle, \label{eq:21} \\
\chi_0^B &=& \langle \Phi_{\bfr_B} \Phi_{\bfr_B} \rangle, \label{eq:22} \\
\chi_i^A &=& \langle \Phi_{\bfr_B - \bfe_\mu} \Phi_{\bfr_B - \bfe_\nu} \rangle, 
\hspace{0.3cm} \bfe_\mu-\bfe_\nu \in jk \text{  plane},  \label{eq:23}\\
\chi_i^B &=& \langle \Phi_{\bfr_A + \bfe_\mu} \Phi_{\bfr_A + \bfe_\nu} \rangle, 
\hspace{0.3cm} \bfe_\mu - \bfe_\nu \in jk \text{  plane}, \label{eq:24} \\
\xi_\mu &=&\langle \Phi^*_{\bfr - e_\mu} \Phi_{\bfr } \rangle. \label{eq:25}
\end{eqnarray}
As mentioned in the previous section, the gauge sector is trivially
soluble, leading to $\Delta = 1/2$.
The spinon action part can be rewritten in matrix notation:
\begin{eqnarray}
S_\Phi =  \int  \frac{d \omega_n}{2\pi} \sum_{\bfk >0}
\vec{\Phi}_{\bfk}^\dagger
\left( M({\bf k}) + \frac{\omega_n^2}{2 J_{zz}} I \right)
\vec{\Phi}_{\bfk} ,
\label{eq:26}
\end{eqnarray}
where
\begin{eqnarray}
\vec{\Phi}_{\bfk} 
&=& \left( \begin{array}{c}
 \Phi_\bfk^A \\ 
 \Phi_{-\bfk}^{A*} \\
 \Phi_{\bfk}^B \\
 \Phi_{-\bfk}^{B*}
 \end{array} \right)  ,
 \label{eq:27} \\
 M({\bf k}) &=& 
 \left( \begin{array}{cccc}
A_{11}(\bfk)& A_{12}(\bfk) & C(\bfk) & 0 \\
A_{12}^*(\bfk) & A_{11} (\bfk) & 0 & C^*(-\bfk) \\
C^*(\bfk) & 0 & B_{11} (\bfk) & B_{12}( \bfk) \\
0 & C(- \bfk) & B_{12}^* (\bfk) & B_{11} (- \bfk)  
\end{array} \right) ,
\label{eq:28}
\end{eqnarray}
and $A_{\alpha ,\beta}(\bfk) $, $B_{\alpha,\beta} (\bfk) $ and $C (\bfk)$ are defined as 
\begin{eqnarray}
A_{11} (\bfk) &=& B_{11} (\bfk) =
-J_{\pm } \Delta^2 \sum_{\mu \neq \nu} e^{ -i \bfk \cdot ( \bfe_\mu - \bfe_\nu)} \nn ,\\
A_{12} (\bfk) &=& \frac{J_{\pm \pm}\Delta^2 }{2}
\sum_{\mu \neq \nu} ( \gamma_{\mu \nu}\vphantom{*} \chi_{\mu \nu}^B 
+ \gamma_{\mu \nu}\vphantom{*} \chi_0^B e^{i \bfk \cdot ( \bfe_\mu - \bfe_\nu ) } ) \nn ,\\
B_{12} (\bfk) &=& \frac{J_{\pm \pm}\Delta^2 }{2}
\sum_{\mu \neq \nu} ( \gamma_{\mu \nu}^* \chi_{\mu \nu}^A 
+ \gamma_{\mu \nu}^* \chi_0^A e^{-i \bfk \cdot ( \bfe_\mu - \bfe_\nu ) } ) \nn ,\\
C (\bfk) &=&  \frac{J_{\pm \pm}\Delta^2 }{2}
\sum_{\mu \neq \nu} 8 \gamma_{\mu \nu} \xi_\mu e^{i \bfk \cdot \bfe_\nu} .
\label{eq:29}
\end{eqnarray}

Finally, to render the mean field problem solvable, we replace the
constraint $|\Phi_\bfr|=1$ by the ``softened" constraint $\frac{1}{N}
\sum_{\bfr} | \Phi_\bfr |^2 =1$, and implemented the latter by
including a Lagrange multiplier term into the action $S_\Phi$.  

Using this formulation, the mean field Hamiltonian allows one to
calculate $\langle H_{QED} \rangle$ (Eq.\eqref{eq:11}) and minimize
this variational energy.  We found and compared several
self-consistent solutions of the gMFT equations, which are subsets of
the general ansatz given above.  First, we considered two limits
allowing for pairing, or A-B sublattice mixing, but not both:
\begin{eqnarray}
\text{(i)}  &&  \xi_\mu =0,\phantom{h}  \chi_0^{A(B)} \neq 0 , \phantom{h}
\chi_{\mu \nu}^{A (B)} \neq 0, \label{eq:30} \\
\text{(ii)} &&  \xi_\mu \neq 0, \phantom{h}  \chi_0^{A(B)} = \chi_{\mu \nu}^{A(B)}=0 .\label{eq:31}
\end{eqnarray}
While self-consistent solutions may be found for both these cases, we
find that the minimum energy solutions always have either vanishing
pairing/sublattice mixing (i.e. describe the $U(1)$ QSL) or exhibit
spinon condensation.   

However, the both condensed solutions are unnatural, insofar as
once a single $\Phi$ field is condensed, all the expectation values
$\chi_0^{A/B}, \chi_i^{A/B}, \xi_\mu$ would be expected to be
non-zero.  Guided by the above cases, we found a self-consistent
ansatz where all these were allowed to be non-vanishing, with the relations $\chi_0^A = \chi_0^B$, $\sum_{\mu \neq \nu} \gamma_{\mu \nu}\chi_{\mu \nu}^B 
= \sum_{\mu \neq \nu} \gamma_{\mu \nu}^* \chi_{\mu \nu}^A \neq 0 $ and
$\xi_0 = \xi_i = - \xi_j =- \xi_k \neq 0$, for $ \{i , j, k\}$ and
permutation of $\{1,2,3\}$.  This more general ansatz describes both
condensed and uncondensed states, and was found to capture all the
physical minimum energy solutions.

\subsubsection{Spinon condensation}
\label{subsec:spinon-condensation}

In the gMFT scheme used here, Higgs phases in which the single spinon
field is condensed, $\langle \Phi_{\bfr}\rangle \neq 0$, also occur.
This may appear surprising since the single spinon field was not
introduced explicitly as an order parameter -- see
Eqs.~\eqref{eq:13} and \eqref{eq:14}.  Instead, spinon condensation occurs,
as discussed in Ref.~\onlinecite{savary:11}, via the same mechanism as
does Bose-Einstein condensation in the non-interacting Bose gas.  In
particular, when a condensate is present, the Lagrange multiplier
$\lambda$ adjusts itself self-consistently so that the minimum energy
spinon state lies, in the thermodynamic limit, at precisely zero energy.
For large but finite volume, a non-intensive part of the $\lambda$ leads
to and controls the condensate, manifesting itself via off-diagonal long
range order in the spinon Green's function.  This is discussed in more
detail in Appendix~\ref{app:2}.  Captured in this way, spinon
condensation does not introduce any additional self-consistent
variables, and only requires careful treatment of any zero energy modes
and the infinite volume limit.  This in turn means that the above
ans\"atze describe Higgs phases as well, for appropriate values of
parameters.

\subsection{gMFT Phase Diagram}
\label{subsec:gMFT}

\begin{figure}[t]
\vskip0.5cm
  \includegraphics[width=0.48\textwidth]{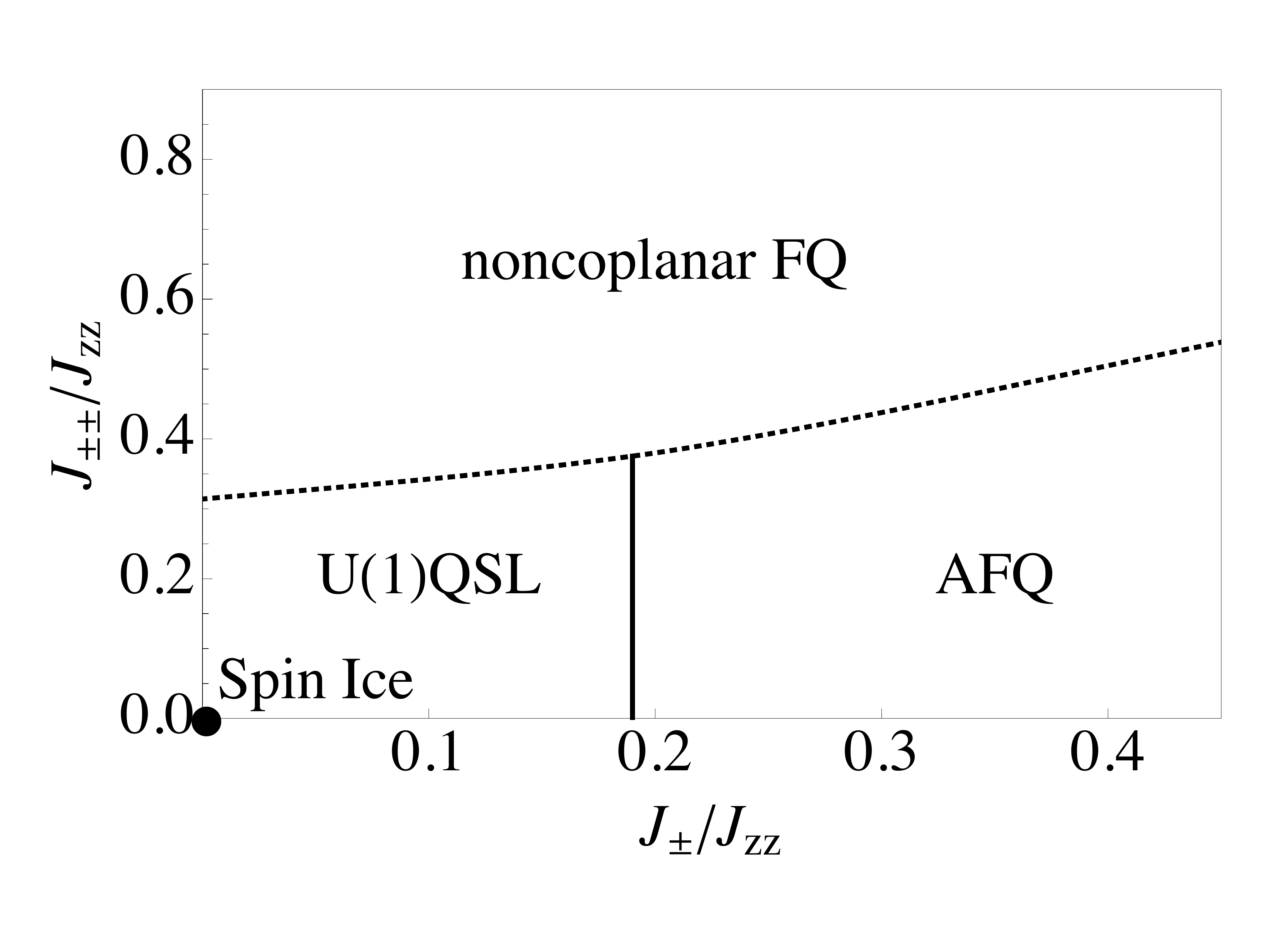}
  \caption{Phase diagram of two dimensionless parameters $J_\pm/J_{zz}$ vs $J_{\pm \pm}/ J_{zz}$. Four distinct phases exist : classical spin ice (at the origin), U(1) QSL, AFQ and FQ. (more details in the main context)  }
\label{fig:3} 
\end{figure}

We minimized the variational energy using the above ansatz numerically
(see Appendix~\ref{app:2} for the formulation of the variational
energy).  In fact, the self-consistent gMFT equations are solved for
any local minima of the variational energy, so it is sufficient to
search for the global minimum of the latter.  That determines the
$T=0$ phase diagram as a function of $J_\pm/J_{zz}>0$ and
$J_{\pm\pm}/J_{zz}$ (we assume $J_{zz}>0$ throughout).  Note that by a
canonical transformation, $\sfS^{\pm} \rightarrow \pm i \sfS^{\pm}$,
we can always choose $J_{\pm\pm}>0$, without loss of generality.  The
results are shown in Fig.\ref{fig:3}.

The full phase diagram contains three distinct phases in addition to the
classical point corresponding to the nearest-neighbor spin ice: a
deconfined U(1) QSL phase and two Higgs phases, corresponding to XY
ferro-pseudospin (antiferro-quadrupolar) and antiferro-pseudospin
(noncoplanar ferro-quadrupolar) orders. Unfortunately, the $Z_2$ spin
liquid phase with non-zero pairing but a spinon gap is never the minimum
energy solution.  The QSL or Coulomb phase occurs in the small $J_\pm,
J_{\pm\pm}$ region, consistent with perturbative expectations.  In this
model, infinitesimal $J_{\pm}$ and/or $J_{\pm \pm}$ interactions
``melt'' the classical spin ice, creating a dynamical ``photon''
excitation and emergent quantum electrodynamics. This phase is found to
be more stable against $J_{\pm\pm}$ than to $J_{z\pm}$, the latter
having been studied already in Ref.~\onlinecite{savary:11}.
%
%
%
\begin{figure}[t]
\vskip0.5cm
  \includegraphics[width=0.5\textwidth]{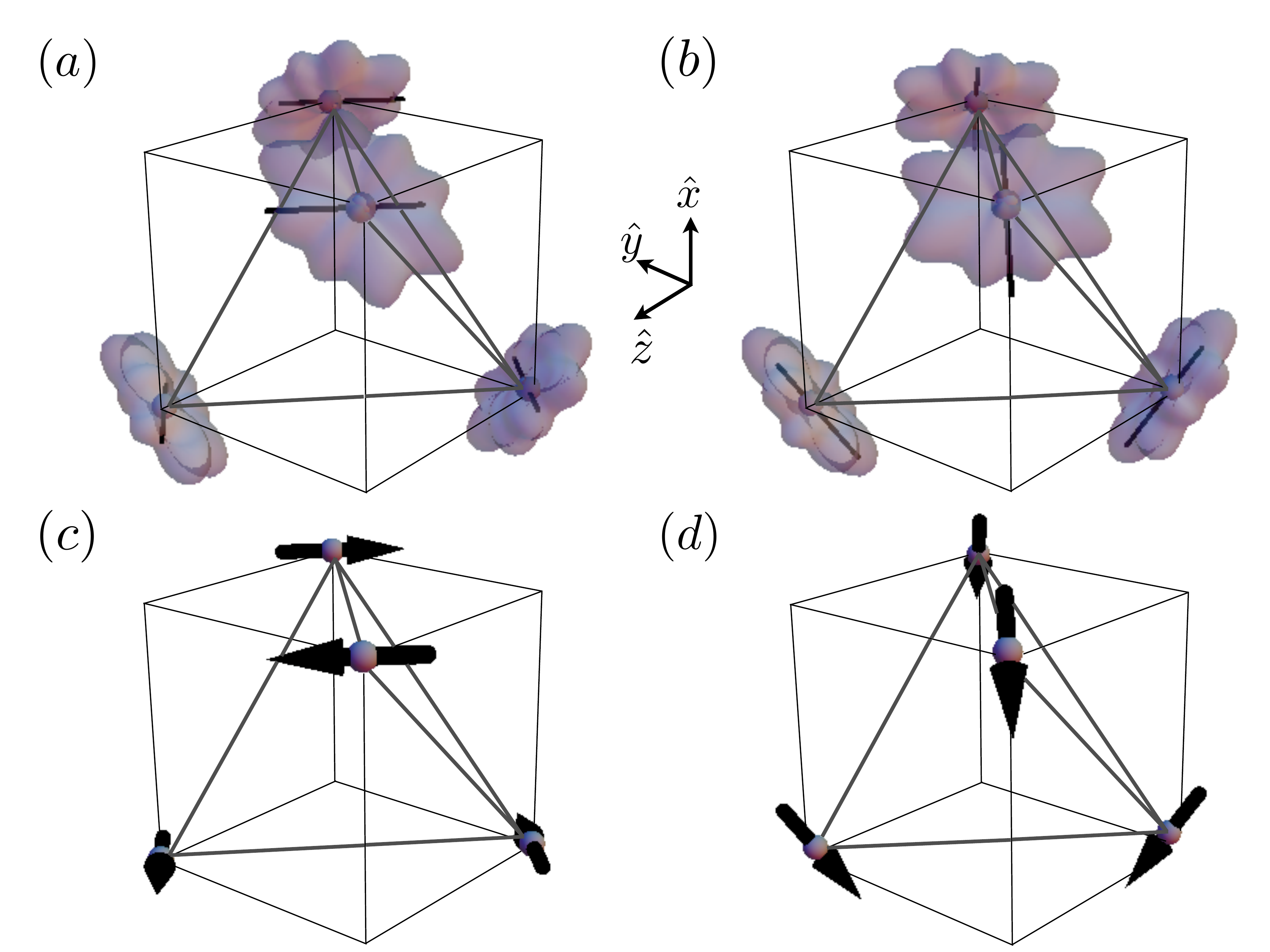}
  \caption{
(a) AFQ :  an $f$-electron charge distribution under the antiferro-quadrupolar order induced by a spinon condensation $\langle \Phi \rangle \neq 0$ at ${\bf k} = (000)$ wave vector. Here, we have taken the Pr$^{3+}$ case with the local ground-state non-Kramers doublet $\alpha| 4\sigma\rangle-\beta\sigma|\sigma\rangle+\gamma|-2\sigma\rangle$ with $\alpha\approx0.970$, $\beta\approx0.075$ and $\gamma\approx0.230$, where $|J_z\rangle$ is an eigenstate of the $z$ component of the total angular momentum in the local frame~\cite{onoda:10}.
(b) Noncoplanar FQ : an $f$-electron charge distribution under the noncoplanar ferroquadrupolar order induced by a spinon condensation $\langle \Phi \rangle \neq 0$ at ${\bf k} = 2\pi (100)$ for the Pr$^{3+}$ case.
(c) Antiferromagnet : an XY antiferromagnetic ordering induced by a spinon condensation $\langle \Phi \rangle \neq 0$ at ${\bf k} = (000)$ wave vector in the case of half-integer spins.
(d) Noncoplanar ferromagnet : an XY noncoplanar ferromagnetic ordering induced by a spinon condensation $\langle \Phi \rangle \neq 0$ at ${\bf k} = 2\pi (100)$ in the case of half-integer spins.
}
\label{fig:4} 
\end{figure}
%

The Higgs or ordered phases merit some further description.  With
increasing $J_{\pm}/J_{zz}$ but $J_{\pm \pm}=0$, the U(1) QSL phase
remains stable untill $\frac{J_{\pm}}{J_{zz}} |_c \approx 0.19$, at
which spinons start to condense at a wave vector $\bfk_0 \equiv 0$ for
both $A$ and $B$ sublattices. This induces a classical XY order
categorized in Table.\ref{tab:1} and has the ordering structure shown in
Fig.\ref{fig:4} (a). This phase has already been obtained by a classical
MF analysis~\cite{onoda:11}, and in gMFT for
$J_{\pm\pm}=0$\cite{savary:11}.  From Eqs.~\eqref{eq:17} and
\eqref{eq:18}, the spinon condensate at $\bfk_0$ yields a ferroic
ordering of the XY component of pseudospins, for instance, given by
\begin{equation}
\langle \vec{\sfS}_i \rangle \approx | \phi_{\bfk_0} |^2  \hat{x}_i,
\label{eq:32}
\end{equation}
for pseudo-spin on sublattice $i$. It spontaneously breaks the threefold rotational symmetry while the twofold rotational symmetries are preserved. This ferro-pseudospin ordering structure is
interpreted as an antiferro-quadrupolar order for Pr$^{3+}$ case as is
clear from Eq.~\eqref{eq:3} and the relation $\sum_{i=0}^3 \hat{x}_i
=0$. 
Namely, it produces an $f$-electron distribution shown in Fig.\ref{fig:4} (a).
When $J_{\pm \pm}>0$ is sufficiently large and $J_{\pm}$ is small, the
QSL becomes unstable to a different Higgs phase, with spinon
condensation at $\tilde{\bfk}_0 \equiv 2\pi(1 00)$ or the symmetry
related points, on both $A$ and $B$ sublattices.  Note that
quantitatively the QSL phase is wider in the $J_{\pm\pm}$ direction than
in the $J_{\pm}$ one: $\frac{J_{\pm\pm}}{J_{zz}}|_c \approx$ 0.31,
compared to $\frac{J_{\pm}}{J_{zz}}|_c \approx 0.19$.  This suggests
that the U(1) QSL phase is more stable against the $J_{\pm \pm}$
interaction than the $J_{\pm}$ interaction. This can be understood from
degenerate perturbation theory.  The $J_{\pm \pm}$ interaction induces a
non-trivial contribution only at the sixth order, ${\sl O}(J_{\pm
  \pm}^6/J_z^5)$, whereas the comparable term is induced already at
the third order in $J_{\pm}$.  As for the other Higgs phase, the ordering
structure is understood again from Eqs.~\eqref{eq:17} and \eqref{eq:18}.
One of the symmetry-broken ground states is
\begin{equation}
( \langle \vec{\sfS}_0 \rangle,\langle \vec{\sfS}_1 \rangle,\langle \vec{\sfS}_2  \rangle,\langle \vec{\sfS}_3 \rangle )
\approx | \phi_{\tilde{\bfk}_0} |^2  ( \hat{y}_0, \hat{y}_1,-\hat{y}_2, -\hat{y}_3).
\label{eq:33}
\end{equation}
It spontaneously breaks both the threefold rotational symmetry and the cubic symmetry, and loses two of the twofold rotational axes. This antiferro-pseudospin structure is
interpreted as a noncoplanar ferro-quadrupolar order in the Pr$^{3+}$ situation, as is clear from Eq.~\eqref{eq:3} and the
relation $(\hat{y}_0 + \hat{y}_1-\hat{y}_2 - \hat{y}_3) \parallel (100)$.
It creates an $f$-electron distribution shown in Fig.\ref{fig:4} (b).
It is worth to note that the above ferrro- and antiferro-pseudospin structures, associated with the antiferro- and noncoplanar ferro-quadrupole orders shown in Fig.\ref{fig:4} (a) and (b) for Pr$^{3+}$ cases, are directly related to XY-magnetic orderings shown in Fig.4 (c) and (d) when we consider half-integer spin rare-earth pyrochlores instead of integer spin case.

\subsection{Phase transitions}
\label{subsec:phase-transitions}

Within gMFT, the phase transition between the U(1) QSL and AFQ state is
second order, as indicated by a continuous change of the MF variables
$\chi_\mu$ from zero to finite values across the phase boundary (solid
line in Fig.\ref{fig:3}).  A low energy continuum action for this
transition is simply an Abelian Higgs theory, with a charged bosonic
matter field (representing the condensing spinons) coupled to a
dynamical gauge field $A$.  When gauge fluctuations beyond the mean field
are included, such transitions are usually driven weakly first
order.\cite{hlr:74}\ It is interesting that, within gMFT, the phase
boundary between QSL and AFQ phases is precisely vertical, as seen in
Fig.\ref{fig:3}. This is because in the QSL phase arbitrarily close to
the phase boundary, both spinon pairing and A-B sublattice mixing is
absent, so that the $J_{\pm \pm}$ interaction gives zero contribution to
the energy.

By contrast with the above case, we find that the QSL to FQ transition
is strongly first order already in gMFT.  This is indicated by the
dotted line in Fig.\ref{fig:3}.  Fluctuation effects will not change
this conclusion. Note that this phase boundary has a positive slope,
i.e. the FQ state is suppressed by increasing $J_\pm$.  This is because
$J_{\pm}$ interaction prefers instead the AFQ state.

\section{AF case ($J_\pm <0$) : $\pi$-flux state}
\label{sec:pi}

A similar analysis can in principle be completed for the case $J_\pm<0$,
for which the XY-pseudospin order is frustrated.  This case, however,
introduces significant new complexities which are beyond the scope of
the present work, and will be discussed in a future publication.  Here,
we confine ourself to the line in the phase diagram $J_{\pm\pm}=0$ in
the AF pseudospin region. 

As discussed in Sec.\ref{sec:zero}, the $J_{\pm}<0$ favors a
$\pi$-flux state, in which all the hexagons carry a flux $\nabla\times
\overline{A}=\pi$ (mod $2\pi$). For calculations, it is necessary to
choose a gauge with a specific assignment of $\overline{A}_{\bfr,\bfr'}$
having $\pi$ flux, as shown in Fig.\ref{fig:2} (b).  In this case, the
unit cell is doubled compared to the case of FM $J_{\pm}$ and contains
four sublattices (comprising 2 diamond sites in each of the 2 magnetic
unit cells ). This gauge field pattern can be represented as $\overline{
A}_{\bfr,\bfr+\bfe_\mu} = \epsilon_\mu {\bf Q} \cdot \bfr $ where
$(\epsilon_0 , \epsilon_1 , \epsilon_2,\epsilon_3 ) = (0 1 1 0)$ and
${\bf Q }= 2\pi (1 0 0)$.  In the QSL state, this leads to $\langle \sfs^{\pm}_{\bfr,\bfr+
  \bfe_\mu} \rangle = \Delta e^{ i\epsilon_\mu {\bf Q} \cdot \bfr}$, with
$\Delta=1/2$. 

In this fixed gauge, we consider the spinon dispersions
for $J_{\pm \pm}=0$.  Within gMFT (see
Eq.~\eqref{eq:14}), the A and B sublattices are decoupled and the spinon
action is
\bw
\begin{eqnarray}
{\sl S}_\Phi &=& \int d \tau \sum_{\bfr \in A,B}  \frac{1}{2J_z} \partial_\tau \Phi^*_\bfr  \partial \Phi_\bfr
+ \lambda^A  \sum_{\bfr \in A} ( | \Phi_\bfr |^2 -1) 
+ \lambda^B  \sum_{\bfr \in B} ( | \Phi_\bfr |^2 -1)   \nn \\
&& + {J_{\pm }} \Big\{ \sum_{\bfr \in A}  \sum_{\mu \neq \nu} ( \Phi_{\bfr +\bfe_\mu}^* \Phi_{\bfr + \bfe_\nu} 
\langle \sfs_{\bfr, \bfr+\bfe_\mu}^- \rangle \langle \sfs^+_{\bfr, \bfr+ \bfe_\nu} \rangle 
 + \sum_{\bfr \in B} \sum_{\mu \neq \nu} (  \Phi_{\bfr- \bfe_\mu}^* \Phi_{\bfr - \bfe_\nu} 
\langle s_{\bfr,\bfr -\bfe_\mu}^+ \rangle \langle s_{\bfr,\bfr-\bfe_\nu} \rangle   \Big\}
\label{eq:34} \\
&=&  \int \frac{d^3 k}{V_{BZ}} \frac{d {\omega}}{2 \pi}  
 \left( \begin{array}{cc}
 \Phi_{\bfk,1}^{A*} & \Phi_{\bfk,2}^{A*} 
 \end{array} \right)
 \left( \begin{array}{cc}
\frac{ \omega^2}{2J_{zz}}  +  \lambda^A +  P_1^A &  - P_2^A - i P_3^A \\
-P_2^A + i P_3^A & \frac{\omega^2}{2J_{zz}} + \lambda^A- P_1^A 
 \end{array} \right)
 \left( \begin{array}{c}
 \Phi_{\bfk,1}^A \\ 
 \Phi_{\bfk,2}^{A} 
 \end{array} \right) + \left(A \rightarrow B \right) .
 \label{eq:35}
 \end{eqnarray}
\ew
Here 
\begin{eqnarray}
P_1^A = 4 J_{\pm} \Delta^2   \cos \frac{k_y }{2}  \cos \frac{k_z}{2}, \nn \\
P_2^A =4 J_{\pm}  \Delta^2   \sin \frac{k_x}{2} \sin \frac{k_y}{2}, \nn \\
P_3^A = 4 J_{\pm} \Delta^2    \cos \frac{k_x}{2} \sin \frac{k_z}{2}, 
\label{eq:36}
\end{eqnarray}
with $P_1^B = P_1^A ( {\bf k} \rightarrow { \bf k} + \pi(111) )$, $P_2^B
= - P_2^A ( {\bf k} \rightarrow { \bf k} + \pi(111))$ and $P_3^B = 
P_3^A ( { \bf k} \rightarrow { \bf k} + \pi(111) )$.
We now seek a QCP between the U(1) $\pi$-flux QSL and a magnetically
ordered phase.  Since the $A$ and $B$ sublattices are decoupled, it is
sufficient to focus on one sublattice, for instance, $A$.  As before, we
adopt a ``softened" constraint $\frac{1}{N} \sum_r \langle
\Phi_{\bfr}^{A*} \Phi_{\bfr}^A \rangle = 1$, which leads, assuming no
spinon condensation, to
\begin{eqnarray}
 \sqrt{ \frac{J_{zz}}{2J_{\pm} } }\frac{1}{\Delta}  \int \frac{d^3 k}{V_{BZ}} 
 \frac{1+ \frac{ \tilde{ \lambda}}{\sqrt{ \tilde{\lambda}^2   - P^2_\bfk} }}{\sqrt{2 \tilde{\lambda} +
2 \sqrt{ \tilde{\lambda}^2 - P^2_\bfk } }} =1.
\label{eq:37}
\end{eqnarray}
Here we defined $ \tilde{\lambda} = \lambda^A /{J_{\pm} \Delta^2} $ and
$P^2_\bfk=\sum_{\alpha=1,2,3}\left[P_{\alpha\bfk}/{ J_{\pm}
    \Delta^2}\right]^2$.  The spinon condensation point occurs when 
the integrand diverges, which gives $\tilde{\lambda}_c = \sqrt{
  \text{max}_\bfk P^2_\bfk} =4$.  By substituting $\tilde{\lambda}_c =
4$ and evaluating the integration in Eq.~\eqref{eq:35} at this point, we
obtain
\begin{equation}
 \frac{|J_{\pm} | }{J_{zz}} \Big|_c \approx 4.13.
 \label{eq:38}
\end{equation}
This is the main result of this Section.  We observe that the QSL phase
is {\em much} more stable to antiferromagnet $J_\pm$ than to
ferromagnetic $J_\pm$.  This is rather natural since the competing XY
pseudospin order is frustrated in the antiferromagnetic case.  We can
understand this more analytically from the spinon dispersion in the
$\pi$-flux state, which has the form $E^{\alpha}_\bfk = \sqrt{ (
  P_{1\bfk}^{\alpha})^2 + (P_{2\bfk}^{\alpha})^2 +
  (P_{3\bfk}^{\alpha})^2  }$.  This form, which describes states in
either A and B sublattice, has a degenerate set of energy minima,
consisting of lines in reciprocal space (e.g. $E^A_{\bfk}$ is minimized
for $\bfk = (k,0,0)$ with an arbitrary real number $k$, and there are
several other similar minimum energy lines).  In contrast, in the FM
case, $\bfk=(000)$ uniquely gives the minimum energy.  This effectively
lowers the spatial dimensionality at low energies, increasing the
stability of the U(1) QSL.  We note, however, that this line degeneracy
is emergent and is not protected by any symmetry.  Effects beyond gMFT
should be taken into account to further split this degeneracy.  Such
effects would be essential in determining the nature of quadrupolar
ordering in the Higgs phase beyond the critical point. We expect this
physics to lead to a significantly richer phase diagram when $J_{\pm
  \pm}$ interaction is included.

\section{Summary}
\label{sec:summary}

In this paper, we have studied the generic pseudospin-1/2 model
describing nearest-neighbor coupling of ground state magnetic doublets
of rare earth ions on the pyrochlore lattice.  We showed how to extend
the gMFT treatment of Ref.~\onlinecite{savary:11} to take into account
all the symmetry allowed interactions, which requires a significant
extension for the method.  We focused on the case of a non-Kramers
ion, for which three interactions exist: an Ising spin-ice interaction
$J_{zz}$ (which we presume always takes the non-trivial frustrated
sign), a U(1) symmetric planar exchange $J_\pm$ and an in-plane
anisotropic exchange $J_{\pm\pm}$.  For the case of ferromagnetic
symmetric planar exchange, we obtained a complete gMFT solution.  This
situation favors ``zero flux'' states in the gauge theory formulation.
We obtained a finite region in the phase diagram supporting a U(1)
quantum spin liquid (QSL) state, described as a type of emergent
quantum electrodynamics.  Phase transitions from the U(1) QSL state to
two types of planar pseudospin orders were found to occur by the Higgs
mechanism with increasing $J_{\pm}$ and $J_{\pm \pm}$.   For large
$J_\pm$ this yields an antiferro-quadrupolar phase, while large
$J_{\pm\pm}$ yields a ferro-quadrupolar state.  In the case of
antiferromagnetic symmetric planar exchange, a $\pi$-flux state is
preferred in the gauge theory, and the general solution was too
complex to attempt here.  However, we did prove that the increased
frustration in this regime greatly increases the stability of the U(1)
QSL state.  

It is hoped that these results form some basis for understanding
experiments in the non-Kramer's pyrochlores Pr$_2TM_2$O$_7$ with
$TM$=Sn, Ir, and Zr.  Future studies should complete the full phase
diagram in the antiferromagnetic planar symmetric exchange case, and
address the accuracy of the gMFT results by comparison with other
methods, considering the role of further neighbor interactions,
lattice distortions, and disorder.  

 \begin{acknowledgments}
   We thank Hyejin Ju, Zheng-Cheng Gu, EG Moon, Cenke Xu and Lucile Savary for useful discussions.
   S.B. Lee and L. Balents were supported by the DOE through Basic
   Energy Sciences grant DE-FG02-08ER46524.  LB’s research facilities
   at the KITP were supported by the National Science Foundation grant
   NSF PHY-0551164. SO was partially supported by Grants-in-Aid for Scientific Research, Grant No. 19052006 from the Ministry of Education, Culture, Sports, Science, and Technology (MEXT) of Japan and under No. 21740275 and No. 24740253 from the Japan Society of Promotion of Science
 \end{acknowledgments}
 
\appendix

\section{Mean-field approximation}
\label{app:1} 

In this section, we proceed two steps of mean-field (MF) approximation to make Eq.~\eqref{eq:11} soluble.

\subsection{Green's functions and energy} 

\subsubsection{Green's function} 

Using the MF ansatz listed in Eqs.~\eqref{eq:20}-\eqref{eq:25}, the Hamiltonian of particle part are written as 
\bw
\begin{eqnarray}
{H_p} &=& \frac{J_{zz}}{2} \sum_{\bfr } Q_r^2 + \tilde{H}_p  \label{aeq:1} \\
\tilde{H}_p  &=&
- J_{\pm} \Delta^2 \sum_{\bfr } \sum_{ \mu \neq \nu}  \Phi^\dagger_{\bfr+ \eta_\bfr\bfe_\mu} \Phi_{\bfr+ \eta_\bfr\bfe_\nu} 
\nonumber\\
&&+ \frac{J_{\pm \pm} \Delta^2 }{2} \sum_{\bfr  } \left\{ \sum_{\mu \neq \nu} \gamma_{\mu \nu}^{-2\eta_\bfr} \left(\chi^{\Gamma_{\eta_\bfr}}_0\right)^*  \Phi_{\bfr+ \eta_\bfr\bfe_\mu} \Phi_{\bfr +\eta_\bfr\bfe_\nu} 
+  \sum_{\alpha =1}^{3} 4 \gamma^{-2\eta_\bfr}_\alpha\chi_\alpha^{\Gamma_{\eta_\bfr}}
\Phi^\dagger_{\bfr} \Phi^\dagger_{ \bfr} 
+ h.c \right\} \nn \\
&& + 2 J_{\pm \pm} \Delta^2 \sum_{\bfr} \left\{  \sum_{\mu \neq \nu} 
\gamma_{\mu \nu}^{ -2 \eta_\bfr}  
\xi_{\mu} \Phi_{\bfr}^\dagger \Phi_{\bfr + \eta_\bfr \bfe_\nu } 
 +h.c  \right\}
\label{aeq:2}
\end{eqnarray}
\ew
with $\Gamma_+=B$ and $\Gamma_-=A$, which leads to the action,
\begin{eqnarray}
{\sl S}_p 
&=& \int d \tau \sum_{\bfr \in A,B}   \frac{1}{2J_{zz}} \partial_\tau \Phi_\bfr^\dagger \partial_\tau \Phi_\bfr 
 + \tilde{H_p} 
 \nonumber\\
 &&+\sum_{\bfr \in A,B} \lambda_\bfr  ( | \Phi_\bfr |^2 -1) )
\label{aeq:3}
\end{eqnarray}
In Eq.~\eqref{aeq:3}, the first term comes from integrating out $Q_\bfr$ and the last term is for Lagrange multiplier which constrains $ | \Phi_\bfr |^2 =1$. In large $N$ limit, the integrals become sharply peaked at the saddle point, say $\lambda^{A(B)}$. Hence we pull out $\lambda_r$ from the summation with its saddle point value $\lambda^{A(B)}$. This is consistent with softening local constraint $ | \Phi |^2 =1$ to its average $\sum_\bfr | \Phi_\bfr |^2 =N$.  Using this saddle point approximation, Eq.~\eqref{aeq:3} can be rewritten in a Fourier transform and this results in Eq.~\eqref{eq:26}. Fourier transform of $\Phi_{\bfr,\tau}$ is defined,
 \begin{equation}
 \Phi_{\bfr, \tau} =   \frac{1}{N_{u.c}} \int \frac{d \omega}{2\pi} \sum_{\bf k} \Phi_{ \bfk,\omega} e^{- i (\omega \tau - \bfk \cdot \bfr)}
 \label{aeq:4}
 \end{equation}
 where $\bfk$ is wave vector, $\omega$ is an imaginary frequency and $N_{u.c}$ is the number of unit cell. 
Then Green's functions are represented as
\begin{eqnarray}
{\mathcal G}_{\alpha \beta} ( \bfk , \omega) \equiv 
\langle  \Phi_\alpha^* \Phi_\beta \rangle 
=\sum_m \frac{2 J_{zz} \phi_\alpha^{m*} \phi_\beta^m }{ \omega^2 +2 J_{zz} ( \lambda + \epsilon_m) }
\end{eqnarray}
The Matsubara sum of frequency leads
\begin{eqnarray}
{\mathcal G}_{\alpha \beta} ( \bfk ) &=& \int \frac{d \omega}{2 \pi} {\mathcal G}_{\alpha \beta} ( \bfk , \omega )  \nn \\
&=& \sum_m \frac{2J_{zz}  \phi_\alpha^{m *} \phi_\beta^m }{ 2 \sqrt{2 J_{zz} ( \lambda + \epsilon_m )}}
 = \sum_m \frac{J_{zz}}{\omega_m} \phi_\alpha^{m*} \phi_\beta^m  \nn 
 \\
\end{eqnarray} 
where $\phi_\alpha^m$ is the $\alpha$ th component of $m$ th eigenvectors for $M$, $\epsilon_m$ is the $m$ th eigenvalues for $M$ and $\omega_m = \sqrt{2 J_{zz} (\lambda + \epsilon_m)}$.

\subsubsection{Variational energy}
\label{app:2}
We consider the energy $\langle H_{QED} \rangle $ by taking an expectation value of Eq.~\eqref{eq:11}.
\bw
\begin{eqnarray}
\langle H_{QED}  \rangle &=& 
  \frac{J_{zz}}{2} \sum_{\bfr} \langle Q_\bfr^2 \rangle 
 -{J_{\pm}} \sum_{\bfr} \sum_{ \mu \neq \nu}  \langle \Phi_{\bfr+ \eta_\bfr \bfe_\mu}^\dagger  \Phi_{ \bfr+ \eta_\bfr \bfe_\nu} \rangle \langle  \sfs_{\bfr,\bfr+ \eta_\bfr \bfe_\mu}^{-\eta_\bfr}  \rangle \langle \sfs_{\bfr, \bfr+ \eta_\bfr \bfe_\nu}^{+\eta_\bfr} \rangle 
 \nn \\
&&  + \frac{J_{\pm \pm}}{2}  \sum_{\bfr} \sum_{\mu \neq \nu}  
\Big\{ \gamma_{\mu \nu}^{-2\eta_\bfr} \Big( \langle \Phi_{\bfr }^\dagger \Phi_{\bfr }^\dagger \rangle 
 \langle \Phi_{\bfr + \eta_\bfr \bfe_\mu} \Phi_{\bfr + \eta_\bfr \bfe_\nu}  \rangle 
 + 2 \langle    \Phi_{\bfr }^\dagger  \Phi_{\bfr + \eta_\bfr \bfe_\mu}  \rangle
\langle \Phi_{\bfr }^\dagger  \Phi_{\bfr + \eta_\bfr \bfe_\nu}  \rangle \Big) 
 \langle  \sfs_{\bfr , \bfr+ \eta_\bfr \bfe_\mu}^{\eta_\bfr} \rangle 
 \langle \sfs^{\eta_\bfr}_{\bfr, \bfr+ \eta_\bfr \bfe_\nu}  \rangle + c.c )\Big\} \nn 
 \\
\label{aeq:7}
\end{eqnarray}
\ew
First of all, let's consider $J_{zz} /2 \sum_{\bfr } \langle Q_\bfr^2 \rangle$. As we mentioned in Sec.\ref{subsec:self-con}, this term can be represented as
\begin{widetext}
\begin{eqnarray}
\frac{J_{zz}}{2} \sum_{\bfr} \langle Q_\bfr^2 \rangle &=& \frac{J_{zz}}{2} \sum_{\bfr } \langle p_{x_\bfr}^2 + p_{y_\bfr}^2 \rangle
\label{aeq:8} \\
&=&   \int \frac{d \omega}{2\pi} \sum_{\bfk \in A,B} ( 1 - \frac{1}{2 J_{zz} } 
\omega^2 \langle \Phi_{\bfk} \Phi_{\bfk} \rangle ) \label{aeq:9} \\
&=&  \int \frac{d \omega}{2 \pi} \sum_{\bfk} \Big( 2 - \frac{1}{2J_{zz}} 
\omega^2 ( {\mathcal G}_{11}  ( \bfk, \omega) + {\mathcal G}_{33}  (\bfk, \omega)) \Big)  \label{aeq:10} \\ 
&=&  \sum_{\bfk} \sum_m \frac{1}{2} \omega_m ( \phi_1^{m *} \phi_1^m + \phi_3^{m*} \phi_3^m )
\label{aeq:11} 
\end{eqnarray}
\end{widetext}
 Here, we used $ \langle p^2 \rangle = 1/ {\sl Z} \int dp dx  p^2 e^{- \int d \tau 
( p^2 + i p \dot{x} +f(x) ) } = 1- \langle \dot{x}^2 \rangle  $ where $Z$ is partition function. 

Finally, Eq.~\eqref{aeq:7} can be rewritten as,
\bw
\begin{eqnarray}
\langle H_{QED} \rangle &=& 
\sum_\bfk \sum_m \frac{1}{2} \omega_m
( \phi_1^{m*} \phi_1^m + \phi_3^{m*} \phi_3^m )
+  \sum_\bfk \sum_m \frac{2J_{zz}}{2\omega_m}
 (A_{11} \phi_1^{m*} \phi_1^m + A_{11} \phi_3^{m*} \phi_3^m ) \nn \\
&& +
 \frac{J_{\pm \pm} \Delta^2 }{2 N_{u.c}} \Big[ 
\Big\{  \sum_{\bfk_A} \sum_m \frac{2J_{zz} }{2 \omega_m} \phi_1^{m*} \phi_2^{m}  \Big\} 
\Big\{  \sum_{\bfk_B} \sum_{m'} \frac{2J_{zz}}{2 \omega_{m'}} \phi_4^{m*'} \phi_3^{m'} 
\sum_{\mu \neq \nu} \gamma_{\mu \nu} e^{- i \bfk_B  \cdot (\bfe_\mu -\bfe_\nu)}\Big\} \nn \\
&&\phantom{hellohel} +\Big\{  
 \sum_{\bfk_A} \sum_m \frac{2J_{zz} }{2 \omega_m} \phi_1^{m*} \phi_2^{m}  
\sum_{\mu \neq \nu} \gamma_{\mu \nu} e^{ i \bfk_A  \cdot (\bfe_\mu -\bfe_\nu)} \Big\} 
\Big\{  \sum_{\bfk_B} \sum_{m'} \frac{2J_{zz}}{2 \omega_{m'}} \phi_4^{m*'} \phi_3^{m'}  \Big\}
+c.c \Big] \nn \\
&& + 
 \frac{J_{\pm \pm} \Delta^2 }{2 N_{u.c}}
 \Big[ \sum_{\mu \neq \nu} 4 \gamma_{\mu \nu} 
 \Big\{\sum_{\bfk_A} \sum_m 
 \frac{2J_{zz}}{2 \omega_m} \phi_1^{m*} \phi_3^m e^{i \bfk_A \cdot \bfe_\mu} \Big\}
\Big\{  \sum_{\bfk_B} \sum_{m'} \frac{2J_{zz}}{2 \omega_m} \phi_1^{m*'} \phi_3^{m'} e^{i \bfk_B \cdot \bfe_\nu}  \Big\} +c.c \Big]
\label{aeq:12}
\end{eqnarray}
\ew

\subsection{spinon condensation}
\label{sec:spinon-cond}
When spinons condense at momentum $\bfk_0$, summation of $\bfk$ in the first Brillouin zone 
can be replaced by 
\begin{eqnarray}
\frac{1}{N_{u.c}} \sum_{\bfk} g( \bfk )  \rightarrow  \frac{ g( \bfk_0) }{N_{u.c}} 
+ \frac{1}{N_{u.c} } \sum_{\bfk \neq \bfk_0 } g( \bfk) 
\label{aeq:13}
\end{eqnarray}
Spinon condensation also affects to a lagrangian multiplier term
and leads $\lambda$ to be
\begin{eqnarray}
\lambda = \lambda_0 + \frac{\lambda'}{{N_{u.c}}^2}
\label{aeq:14}
\end{eqnarray}
where $\lambda_0$ is the minimum of $\epsilon_m$. 
\bw
\begin{eqnarray}
\frac{1}{N_{u.c}} \sum_{\bfk } \sum_m f \Big( \bfk, \omega_m (\lambda, \bfk) \Big) \phi_\alpha^{m*} ( \bfk) \phi_\beta^m ( \bfk )  \rightarrow &&
f \Big( \bfk_0, \omega_{\bar{m}} ( {\lambda'}, \bfk_0) \Big)
 \phi_\alpha^{ \bar{m} * } (\bfk_0) \phi_\beta^{\bar{m} }  (\bfk_0) \nn  \\
 && + \int_{\bfk \neq \bfk_0} \frac{d^3 k}{V_{BZ}} \sum_m f \Big( \bfk , \omega_m (\lambda_0 , \bfk) \Big)
\phi_\alpha^{m*} (\bfk) \phi_\beta^m (\bfk)
\label{aeq:15}
\end{eqnarray}
\ew
$\bar{m}$ is the $m$ th eigenvectors of $M$ which has the minimum eigenvalue of $\epsilon_m$.


\end{document}